\documentclass[twocolumn,showpacs,aps,prl,superscriptaddress]{revtex4}
\usepackage{graphicx,amsmath}
\usepackage{dcolumn}

\newcommand{\bzjpsipiz}{\ensuremath{\Bz\to \jpsi\piz}}
\newcommand{\bzjpsiks}{\ensuremath{\Bz\to\jpsi\KS}}
\newcommand{\bzjpsikspizpiz}{\ensuremath{\Bz\to\jpsi\KS(\piz\piz)}}
\newcommand{\De} {$\mathrm{\Delta}E$}
\newcommand{\DT} {\mbox{$\Delta t$}\xspace}

\newcommand{\jpsifake} {\ensuremath{\jpsi_{\rm fake}}}

\newcommand{\BABARPubYear}    {03}
\newcommand{\BABARPubNumber}  {003}

\newcommand{\SLACPubNumber} {9668}

\input pubboard/babarsym

\def\figurebox#1#2#3{%
    \def\arg{#3}%
    \ifx\arg\empty
    {\hfill\vbox{\hsize#2\hrule\hbox to #2{\vrule\hfill\vbox to #1{\hsize#2\vfill}\vrule}\hrule}\hfill}%
    \else
    {\hfill\epsfbox{#3}\hfill}%
    \fi}

\begin{document}

\preprint{\babar-PUB-\BABARPubYear/\BABARPubNumber}
\preprint{SLAC-PUB-\SLACPubNumber} 

\begin{flushleft}
\babar-PUB-\BABARPubYear/\BABARPubNumber \\
SLAC-PUB-\SLACPubNumber\\
%hep-ex/\LANLNumber\\[10mm]
\end{flushleft}

\title{
{\large \bf
Study of Time-Dependent {\boldmath ${C\!P}$} Asymmetry in Neutral
{\boldmath $B$} Decays to 
{\boldmath ${J\mskip -3mu/\mskip -2mu\psi\mskip 2mu \pi^0}$}}
}

% Author list
%% author list as of 01-Jan-2003 (551 authors)
%
\author{B.~Aubert}
\author{R.~Barate}
\author{D.~Boutigny}
\author{J.-M.~Gaillard}
\author{A.~Hicheur}
\author{Y.~Karyotakis}
\author{J.~P.~Lees}
\author{P.~Robbe}
\author{V.~Tisserand}
\author{A.~Zghiche}
\affiliation{Laboratoire de Physique des Particules, F-74941 Annecy-le-Vieux, France }
% ...
\collaboration{The \babar\ Collaboration}
\noaffiliation

\date{\today}% It is always \today, today, but you may specify any date with \date.

\begin{abstract}
  We present the first study of the time-dependent ${C\!P}$-violating
  asymmetry in ${B^0\rightarrow J\mskip -3mu/\mskip -2mu\psi\mskip
  2mu \pi^0}$ decays using ${e^+e^-}$ annihilation data collected with the
  ${\mbox{\sl B\hspace{-0.4em} {\small\sl A}\hspace{-0.37em} 
  \sl B\hspace{-0.4em} {\small\sl A\hspace{-0.02em}R}}}$ detector at the
  ${\Upsilon{\rm( 4S)}}$ resonance during the years 1999--2002 at the
  PEP-II asymmetric-energy $B$ Factory at SLAC.
  Using approximately $88$ million \BB\ pairs, our results for
  the coefficients of the cosine and sine
  terms of the ${C\!P}$ asymmetry are $C_{\jpsi\piz} = 0.38 \pm 0.41\
  \stat \pm 0.09\ \syst$ and $S_{\jpsi\piz} = 0.05 \pm 0.49\ \stat
  \pm 0.16\ \syst$.
\end{abstract}

\pacs{13.25.Hw, 12.15.Hh, 11.30.Er}% PACS, the Physics and Astronomy Classification Scheme.

\maketitle
% --- The body of the paper starts here ------------------------------------------
% Introduction
The Standard Model of electroweak interactions describes
${C\!P}$ violation in \B-meson decays by a complex phase in the
three-generation Cabibbo-Kobayashi-Maskawa (CKM) quark-mixing
matrix~\cite{ref:CKM}.  The ${b \rightarrow c\mskip 2mu \overline c \mskip
2mu s}$ modes such as
${B^0 \rightarrow J\mskip -3mu/\mskip -2mu\psi\mskip 2mu 
K^0_{\scriptscriptstyle S}}$ yield precise measurements of
the quantity $\stwob$, where $\beta \equiv \arg \left[\,
  -V_{cd}^{}V_{cb}^* / V_{td}^{}V_{tb}^*\, \right]$
(see for example Refs.~\cite{ref:BABARs2b, ref:babar_s2b_new_prl, ref:BELLEs2b}).
The decay \bzjpsipiz\ is a Cabibbo-suppressed
${b \rightarrow c\mskip 2mu \overline c \mskip 2mu d}$
transition.
In the Standard Model both \bzjpsiks\ and \bzjpsipiz\ have penguin
amplitudes with the same weak phase as the tree amplitude, and an
additional penguin amplitude with a different phase.
In \bzjpsiks, the penguin amplitude with a different weak phase
is suppressed by $\lambda_{CKM}^2$, where $\lambda_{CKM}$ is the sine
of the Cabibbo angle, while in \bzjpsipiz, the tree and each penguin
amplitude are equal to leading order in $\lambda_{CKM}$.
Therefore, \bzjpsipiz\ may have a ${C\!P}$ asymmetry that differs
from that of \bzjpsiks, with the size of the asymmetry serving as a
probe of the penguin decay amplitudes in both modes.

\babar\ has previously measured the \bzjpsipiz\ branching fraction,
$(2.0 \pm 0.6\ \stat \pm 0.2\ \syst) \times 10^{-5}$~\cite{ref:excl_B_PRD},
using \FourS\to\BB\ decays.  For the ${C\!P}$ asymmetry measurement,
the flavor (\Bz\ or \Bzb) of the \B\ meson that decays to $\jpsi\piz$
is inferred, or tagged, using properties of the other \B\ meson and the
time evolution of the \BB\ system.  The decay time distributions,
${\rm f}_+({\rm f}_-)$, of \B\ decays to a ${C\!P}$ eigenstate
with a \Bz (\Bzb) flavor tag, are given by
\begin{eqnarray}
{\rm f}_\pm(\, \deltat) = {\frac{e^{{- \left| \deltat \right|}/\tau_{\Bz}
}}{4\tau_{\Bz} }}  
\Bigg[ 1 \Bigg.& \!\!\! \pm& \!\!\! S_{J/\psi\pi^0}
  \sin{( \Delta m_{d}  \deltat )} \nonumber \\
 &\!\!\! \mp& \!\!\! \Bigg.C_{J/\psi\pi^0}
       \cos{( \Delta m_{d}  \deltat) }  \Bigg],
\label{eq:timedist}
\end{eqnarray}
where $\Delta t = t_{\rm rec} - t_{\rm tag}$ is the difference between
the proper decay time of the reconstructed $B$ meson and 
the proper decay time of the tagging $B$ meson,
$\tau_{\Bz}$ is the \Bz\ lifetime, and \deltamd is the \Bz-\Bzb\ oscillation
frequency.  The coefficients can be expressed in terms of a complex
parameter $\lambda$, which depends on both the \Bz-\Bzb oscillation
amplitude and the \Bz\ and \Bzb\ decay amplitudes to this final
state~\cite{ref:lambda}:
$S_{J/\psi\pi^0} = {2\mathop{\cal I\mkern -2.0mu\mit m}}
\lambda / (1+|\lambda|^2)$ and
$C_{J/\psi\pi^0} = (1  - |\lambda|^2) / (1+|\lambda|^2)$.
A decay amplitude
with only a tree component would give $S_{\jpsi\piz} = -\stwob$ and
$C_{\jpsi\piz} = 0$.

% The BaBar detector and dataset
The data used in this measurement were collected with the
\babar\ detector~\cite{ref:babar} at the \pep2\ storage ring in the
years 1999 to 2002.  Approximately $81\invfb$ of ${e^+e^-}$
annihilation data recorded at the \FourS\ resonance are used,
corresponding to a sample of approximately $88$ million \BB\ pairs.
An additional $5\invfb$ of data collected approximately $40 \mev$
below the \FourS\ resonance are used to characterize
non-\BB\ background sources.

% Candidate selection
\bzjpsipiz\ candidates are selected (details are given in
Ref.~\cite{ref:excl_B_PRD}) by identifying \jpsi\to\epem\ or
\jpsi\to\mumu\ decays and \piz\to\gaga\ decays.  For the
\jpsi\to\epem\ (\jpsi\to\mumu) channel, each lepton candidate must be
consistent with the electron (muon) hypothesis.  The invariant mass of
the lepton pair is required to be between $2.95$ and $3.14 \gevcc$,
and $3.06$ and $3.14 \gevcc$, for the electron and muon channels,
respectively.  The photon candidates used to reconstruct the
\piz\ candidate are identified as clusters in the electromagnetic
calorimeter (EMC) with polar angles between $0.410$ and $2.409 \rad$,
that are spatially separated from every charged track, and have a
minimum energy of $30 \mev$.  The lateral energy distribution in the
cluster is required to be consistent with that of a photon.  The
invariant mass of the photon pair is required to between $100$ and
$160 \mevcc$.  Finally, the \jpsi\ and \piz\ candidates are assigned
their nominal masses and combined using 4-momentum addition.

Two kinematic consistency requirements are applied to each \B\ candidate.
The difference, \De, between the \B-candidate energy and the
beam energy in the \epem\ center-of-mass (CM) frame must be
$-0.4 < \mathrm{\Delta}E < 0.4 \gev$.  The beam-energy-substituted
mass, $\mes=\sqrt{(\sqrt{s}/2)^2-(p_B^{\rm *})^2}$, must
be greater than $5.2 \gevcc$, where $\sqrt{s}$ is the total CM energy
and $p_B^{\rm *}$ is the \B-candidate momentum in the CM frame.

A linear combination of several kinematic and topological variables,
determined with a Fisher discriminant, provides 
additional separation between signal and
\epem\to\qqbar ($\q = \u,\d,\s,\c$) continuum background events.
The Fisher discriminant uses the following inputs:  the zeroth- and second-order
Legendre polynomial momentum moments ($L_0 = \sum_i |{\bf p}^{\rm *}_i|$ and
$L_2 = \sum_i |{\bf p}^{\rm *}_i| \hspace{0.5mm} \frac{3 \cos^2\theta_i - 1}{2}$,
where ${\bf p}^{\rm *}_i$ are the CM momenta for the tracks and neutral
calorimeter clusters that are not associated with the signal candidate, and
$\theta_i$ are the angles between ${\bf p}^{\rm *}_i$ and the thrust axis of
the signal candidate);  the ratio of the
second-order to zeroth-order Fox-Wolfram moments, again using just
tracks and clusters not associated with the signal candidate;
$|\cos{\theta_T}|$, where $\theta_T$ is the angle between the thrust
axis of the \B\ candidate and the thrust axis of the
remaining tracks and clusters in the event;
and $|\cos{\theta_\ell}|$, where $\theta_\ell$
is defined as the angle between the
negative lepton and \B\ candidate directions in the \jpsi\ rest frame.
The requirement placed on the Fisher discriminant is $99\%$
efficient for signal and rejects $71\%$ of the continuum background.
The efficiencies for satisfying this requirement are summarized in
Table~\ref{table:efficiencies}.

\begin{table}
\caption{Efficiencies for the requirement on the Fisher discriminant
  and flavor tagging, given independently, with statistical uncertainties.}
\label{table:efficiencies}
\begin{ruledtabular}
\begin{tabular*}{\hsize}{l
@{\extracolsep{0ptplus1fil}}  D{,}{\ \pm\ }{-1}
@{\extracolsep{0ptplus1fil}}  D{,}{\ \pm\ }{-1}}
Type of event      &
\multicolumn{2}{c}{Efficiency (\%)} \\
                   &
\multicolumn{1}{c}{\,\, Fisher} &
\multicolumn{1}{c}{\,\, Tagging} \\ \colrule
\\[-2.5mm]
\bzjpsipiz\                                & 99.2,0.1 & 65.6,0.6 \\
\bzjpsikspizpiz\ bkg.                      & 98.9,0.1 & 65.6,0.6 \\
Inclusive \jpsi\ bkg.                      & 94.9,0.7 & 70.4,1.4 \\
\BB\ generic bkg.                          & 98.5,0.4 & 61.1,1.6 \\
Continuum bkg.                             & 28.6,0.7 & 52.3,0.8 \\
\end{tabular*}
\end{ruledtabular}
\end{table}

% Backgrounds
We split the backgrounds into four mutually exclusive categories, two
of which have a \jpsi\ from \B\ decays (\B\to\jpsi\!X).  The
first background category is \bzjpsikspizpiz\ decays where one of the
\piz\ mesons is nearly at rest in the \epem\ CM frame.
The second background category consists of other
\B\to\jpsi\!X decays (inclusive \jpsi), which contribute through
random combinations of \jpsi\ and \piz\ candidates.
The third and fourth categories consist of random combinations of
particles in \BB\ decays (\BB\ generic) and continuum events,
respectively.  Monte Carlo simulation~\cite{ref:geant4} is used to
model aspects of the \bzjpsikspizpiz, inclusive \jpsi, and
\BB\ generic backgrounds.  A sample (\jpsifake) selected from data
taken below the \FourS\ resonance is used to model the continuum
background.  In this case, the \jpsi\ candidate is reconstructed from
two tracks that are not consistent with a lepton hypothesis.  Monte
Carlo simulation is used to check that this procedure, which increases
the size of the sample, correctly models the continuum background.

% Flavor tagging and measurement of deltaT
The algorithm for \B-flavor tagging assigns events to one of four
hierarchical, mutually exclusive tagging categories, and is described
in detail in Ref.~\cite{ref:babar_s2b_new_prl}.  The total tagging
efficiency for the signal and each background source is given in
Table~\ref{table:efficiencies}.
Untagged events are excluded from further consideration.
Vertex reconstruction and the determination of \DT follow the
techniques detailed in Ref.~\cite{ref:babar_s2b_PRD}.  We require
$-20 < \Delta t < 20 \ps$ and an estimated uncertainty on \DT of less
than $2.4 \ps$.

% Maximum likelihood fitting technique
We extract the ${C\!P}$ asymmetry by performing an unbinned extended
maximum likelihood fit.  The likelihood is constructed from the
probability density functions (PDFs) for the variables \mes,
\De, and \DT.  The quantity that is maximized is the logarithm of
\begin{equation}
\label{eq:CP_likelihood}
{\cal L} = \frac{e^{-\sum_{j=1}^{5}n_j}}{N!}
\hspace{1mm} \prod_{i=1}^{N} \hspace{1mm} \sum_{j=1}^{5}
\left[ f^{\alpha_i}_j \hspace{1mm} n_j
\hspace{1mm} \prod_d \hspace{1mm} {\cal P}^{\hspace{0.25mm}d}_j \right],
\end{equation}
where $n_j$ is the number of events for each of the five hypotheses (one
signal and four background) and $N$ is the number of input events.
The ${\cal P}^{\hspace{0.25mm}d}_j$ are the one- or two-dimensional
PDFs for variables $d$, for each signal or background type.
The parameters $f^{\alpha_i}_j$ are the tagging
fractions for each of the tagging categories $\alpha_i$
(assigned for each event $i$) and each of
the signal or background types $j$.
For the \bzjpsipiz\ signal and \bzjpsikspizpiz\ background, the values
of $f^{\alpha_i}_j$ are measured with a sample
($B_{\rm flav}$) of neutral \B\ decays to flavor eigenstates
consisting of the channels $D^{(*)-}h^+ (h^+=\pi^+,\rho^+$, and $a_1^+)$ and
$\jpsi\Kstarz (\Kstarz\to\Kp\pim)$~\cite{ref:babar_s2b_new_prl}.
Monte Carlo simulation is used to estimate the $f^{\alpha_i}_j$ values
for the inclusive \jpsi\ and \BB\ generic backgrounds, while the
\jpsifake\ sample is used for the continuum background.

% Probability density functions for m_ES and dE
The signal \mes\ distribution is modeled as the sum of two components.
The first is a modified Gaussian function that, for values less than
the mean, has a width parameter that scales
linearly with the distance from the mean.
The second component, accounting for less than $6\%$ of the
distribution, is a threshold function~\cite{argus}, which is a
phase-space distribution of the form
${\mbox{\mes}} \sqrt{(1 - \frac{{\mbox{\mes}}^2}{E_{\rm beam}^2})}
\hspace{1mm} {\rm exp}(\xi(1 -
\frac{{\mbox{\mes}}^2}{E_{\rm beam}^2}))$, with a kinematic cut-off at
$E_{\rm beam} = 5.289 \gev$ and one free parameter $\xi$.
The signal \De\ distribution is modeled by the sum of a Gaussian core
with an asymmetric power-low tail~\cite{CB} and a second order
polynomial.
The parameters of these PDFs are determined by fitting to a signal Monte
Carlo sample.  The peak position of the \De\ distribution is a free
parameter of the full ${C\!P}$ likelihood fit to allow for EMC energy
scale uncertainties.

The kinematic variables \mes\ and \De\ are correlated in the
\bzjpsikspizpiz\ and inclusive \jpsi\ backgrounds, so two-dimensional
PDFs are employed for these modes.  Variably-binned interpolated
two-dimensional histograms of these variables are constructed from the
relevant Monte Carlo samples.

The \mes\ PDFs for the \BB\ generic and continuum backgrounds are
modeled by the threshold function given above, and the \De\ PDFs for
these two backgrounds are modeled by second order polynomials.  The
parameters for these PDFs are obtained from the \BB\ generic Monte
Carlo sample and the \jpsifake\ sample.

% Probability density functions for deltaT
The PDFs used to describe the \DT\ distributions of the signal
and background sources are each a convolution of a resolution
function ${\cal R}$ and decay time distribution~${\cal D}$:
${\cal P}(\Delta t,\sigma_{\Delta t}) = {\cal R}(\delta
t,\sigma_{\Delta t}) \otimes {\cal D}(\Delta t_{\rm true})$,
where $\Delta t$ and $\Delta t_{\rm true}$ are the measured and true decay
time differences, $\delta t$ = $\Delta t - \Delta t_{\rm true}$,
and $\sigma_{\Delta t}$ is the estimated event-by-event error on \DT.

For the signal, the resolution function consists of the sum of three
Gaussian distributions, the parameters of which are determined from
the $B_{\rm flav}$
sample, as in the \bzjpsiks\ measurement~\cite{ref:babar_s2b_PRD}.
The decay time distribution is given by Eq.~\ref{eq:timedist} modified
for the effects of \B-flavor tagging:
\begin{align}
\label{eq:dt_sig_D}
{\cal D}_{\alpha,f}^{\pm}(\deltat) = & {\frac{{e}^{{- \left| \deltat
        \right|}/\tau_{\Bz} }}{4\tau_{\Bz}}} 
\{ (1 \mp \Delta w_{\alpha}) \nonumber \\
& \pm S_{f} \hspace{1mm} (1 - 2w_{\alpha}) \sin(\deltamd  \deltat)
\nonumber \\
& \mp C_{f} \hspace{1mm} (1 - 2w_{\alpha}) \cos(\deltamd \deltat)
 \},
\end{align}
where ${\cal D}_{\alpha,f}^{+}$(${\cal D}_{\alpha,f}^{-}$) is for a $\Bz$($\Bzb$) tagging meson.
The variable $w_{\alpha}$ is the average probability of
incorrectly tagging a \Bz\ as a \Bzb\ ($w^{\Bz}_{\alpha}$) or
a \Bzb\ as a \Bz ($w^{\Bzb}_{\alpha}$), and 
$\Delta w_{\alpha} = w^{\Bz}_{\alpha} - w^{\Bzb}_{\alpha}$.
Both $w_{\alpha}$ and $\Delta w_{\alpha}$ are determined using the
$B_{\rm flav}$ data sample~\cite{ref:babar_s2b_new_prl}.
We use the values $\deltamd = 0.489 \ps^{-1}$ and
$\tau_{\Bz} = 1.542 \ps$~\cite{2002PDG}.

The PDF used to model the \DT\ distribution for the
\bzjpsikspizpiz\ background, which also includes a $C\!P$ asymmetry,
takes the same form as that for signal, but with
$S_{\jpsi\KS} = \stwob = 0.74$~\cite{ref:babar_s2b_new_prl}
and $C_{\jpsi\KS} = 0$.

The parameterizations of the \DT\ PDFs for the inclusive
\jpsi\ and \BB\ generic backgrounds each consist of prompt and
exponential decay components.  Decays appear to be prompt when
particles from the reconstructed \B\ are erroneously included in the
tagging \B\ vertex.  For the \BB\ generic background, the
prompt and exponential components correspond to the cases where the
two decay products forming the \jpsi\ come from both or just one of
the \B\ mesons, respectively.  The fraction that is in the exponential
component, the decay lifetime parameter, and the resolution parameters
are determined from the Monte Carlo simulation.

The \DT\ PDF for the continuum background has only a prompt
component and the resolution parameter values are obtained by fitting
the \jpsifake\ sample.

\begin{figure}[!ht]
\begin{center}
\includegraphics[width=8.5cm,angle=0]{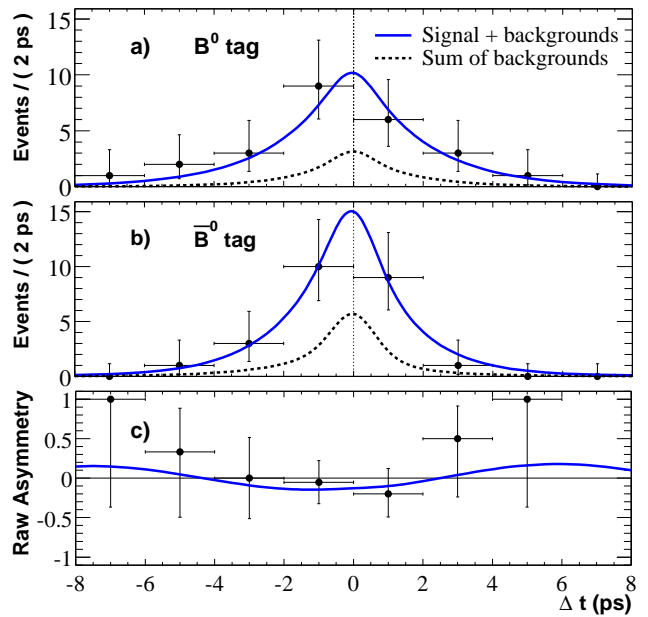}
\caption{Distributions of events a) with a \Bz\ tag
  ($N_{\Bz}$), b) with a \Bzb\ tag ($N_{\Bzb}$), and c) the raw
  asymmetry $(N_{\Bz} - N_{\Bzb})/(N_{\Bz} + N_{\Bzb})$, as functions
  of \DT.  Candidates in these plots are required to satisfy
  $-0.11 < \mathrm{\Delta}E < 0.11 \gev$ and $\mes > 5.27 \gevcc$.
  Of the $49$ signal and background events in this region, $25$ have
  a \Bz\ tag and $24$ have a \Bzb\ tag, with fit background
  contributions of approximately $5$ and $7$ events, respectively.
  The curves are projections that use the values of the
  other variables in the likelihood to determine the contributions to
  the signal and backgrounds.
}
\label{fig:fit_asym}
\end{center}
\end{figure}

% Results of the CP asymmetry fit
The results of the ${C\!P}$ asymmetry fit, for all free parameters,
are shown in Table~\ref{table:fit_results}.  There are $40 \pm 7$
signal events in the total sample of $438$ selected events.
The projection in \mes\ is shown in
Fig.~\ref{fig:fit_results}.  The yields and asymmetry as functions
of \DT, overlaid with projections of the likelihood fit results,
are shown in Fig.~\ref{fig:fit_asym}.
Repeating the fit with the added constraint $C_{\jpsi\piz} = 0$ does
not significantly change the result for $S_{\jpsi\piz}$.

\begin{table}
\caption{Results of the ${C\!P}$ likelihood fit, for
  the full region $-0.4 < \mathrm{\Delta}E < 0.4 \gev$ and
  $\mes > 5.2 \gevcc$.  Errors are statistical only.  The global
  correlation coefficient is $0.14$ for $C_{\jpsi\piz}$ and $0.15$ for
  $S_{\jpsi\piz}$.}
\label{table:fit_results}
\begin{ruledtabular}
\begin{tabular*}{\hsize}{l
@{\extracolsep{0ptplus1fil}}  D{,}{\ \pm\ }{-1}}
                              & \multicolumn{1}{c}{\,\,\, Fit results} \\ \colrule
$C_{\jpsi\piz}$                  & 0.38,0.41  \\
$S_{\jpsi\piz}$                  & 0.05,0.49  \\
Signal \De\ peak position (\mev) & -13.2,7.2\\
\bzjpsipiz\ signal          (events)  & 40,7      \\
\bzjpsikspizpiz\ background (events)  & 140,19    \\
Inclusive \jpsi\ background (events)  & 109,35    \\
\BB\ generic background     (events)  & 52,25     \\
Continuum background        (events)  & 97,22     \\
\end{tabular*}
\end{ruledtabular}
\end{table}

\begin{figure}[!ht]
\begin{center}
\includegraphics[width=8.5cm,angle=0]{mes_plot.epsi}
\caption{Projection in \mes\ for the results
  of the ${C\!P}$ fit, displayed with the added requirement
  $-0.11 < \mathrm{\Delta}E < 0.11 \gev$.  In contrast, the 
  ${C\!P}$ fit uses the full \De\ region.
  In the further restricted region $\mes > 5.27 \gevcc$, there
  are $49$ data events (points), of which about $12$ events are fit as
  background.  Here, \bzjpsikspizpiz\ and
  inclusive \jpsi\ decays contribute to the enhancement in the
  background distribution at large \mes.}
\label{fig:fit_results}
\end{center}
\end{figure}

% Systematic uncertainties
The dominant contributions to the systematic errors in $C_{\jpsi\piz}$
and $S_{\jpsi\piz}$ are summarized in
Table~\ref{table:syst_error}.  The first class of uncertainties are
those obtained by variation of the parameters used in the \mes, \De,
and \DT\ PDFs, where the dominant sources are the uncertainties in the
signal \De\ PDF parameters.
A systematic error to account for a correlation between
the tails of the signal \mes\ and \De\ distributions is obtained by
using a two-dimensional PDF.
Another contribution stems from the impact of EMC energy scale 
uncertainties on the modeling of the \bzjpsikspizpiz\ background.
An additional systematic uncertainty comes from the choice of the
binning of the two-dimensional PDFs for the \bzjpsikspizpiz\ and
inclusive \jpsi\ backgrounds.

\begin{table}
\caption{Summary of systematic uncertainties. \hspace{1.5cm}}
\label{table:syst_error}
\begin{ruledtabular}
\begin{tabular*}{\hsize}{
l@{\extracolsep{0ptplus1fil}} c c}
Source          & \multicolumn{1}{c}{$C_{\jpsi\piz}$} & 
\multicolumn{1}{c}{$S_{\jpsi\piz}$} \\ \colrule
\multicolumn{3}{l}{Parameter variations} \\ \hline
\,\,\,\, \mes\ and \De\ parameters            & $0.05$ & $0.13$ \\
\,\,\,\, Tagging fractions                    & $0.00$ & $0.01$ \\
\,\,\,\, \DT\ parameters                      & $0.03$ & $0.02$ \\ \hline
\multicolumn{3}{l}{Additional systematics} \\ \hline
\\[-3mm]
\,\,\,\, \De--\mes\ correlation in signal     & $0.07$ & $0.08$ \\
\,\,\,\, EMC energy scale \bzjpsikspizpiz\    & $0.01$ & $0.00$ \\
\,\,\,\, Choice of two-D histogram PDFs         & $0.01$ & $0.03$ \\
\,\,\,\, Beam spot, boost/vtx., misalignment  & $0.01$ & $0.01$ \\ \hline \hline
\\[-3mm]
Total systematic uncertainty   & $0.09$ & $0.16$ \\
\end{tabular*}
\end{ruledtabular}
\end{table}

% Summary
In summary, an unbinned extended maximum likelihood fit yields
$40 \pm 7$ signal events and the parameters of
time-dependent CP asymmetry
for the decay \bzjpsipiz:  $C_{\jpsi\piz} = 0.38 \pm
0.41\ \stat \pm 0.09\ \syst$ and
$S_{\jpsi\piz} = 0.05 \pm 0.49\ \stat \pm 0.16\ \syst$.
Within the Standard Model formulation of ${C\!P}$ asymmetries, these
results demonstrate the possibility, with additional integrated
luminosity, of observing penguin contributions in \bzjpsipiz.
Such a measurement may experimentally constrain similar amplitudes in
\bzjpsiks.

% Acknowledgments
We are grateful for the excellent luminosity and machine conditions
provided by our \pep2\ colleagues, 
and for the substantial dedicated effort from
the computing organizations that support \babar.
The collaborating institutions wish to thank 
SLAC for its support and kind hospitality. 
This work is supported by
DOE
and NSF (USA),
NSERC (Canada),
IHEP (China),
CEA and
CNRS-IN2P3
(France),
BMBF and DFG
(Germany),
INFN (Italy),
FOM (The Netherlands),
NFR (Norway),
MIST (Russia), and
PPARC (United Kingdom). 
Individuals have received support from the 
A.~P.~Sloan Foundation, 
Research Corporation,
and Alexander von Humboldt Foundation.

\end{document}